\documentclass[aps,prl,twocolumn,showpacs,superscriptaddress]{revtex4-1}
\usepackage{graphicx}
\usepackage{amsmath,amssymb}
\usepackage[utf8]{inputenc}
\usepackage{natbib}
\usepackage{hyperref}
\usepackage{color}
\hypersetup{colorlinks=true,citecolor={blue},linkcolor={blue},urlcolor={blue}}
\usepackage{units}
\usepackage{xr}
\begin{document}

\title{Polariton Condensation in an optically induced 2D potential}

\author{A. Askitopoulos}
\email[correspondence address: ]{A.Askitopoulos@soton.ac.uk}
\affiliation{School of Physics and Astronomy, University of Southampton, Southampton, SO17
 1BJ, United Kingdom}
\author{H. Ohadi}
\affiliation{School of Physics and Astronomy, University of Southampton, Southampton, SO17
 1BJ, United Kingdom}
\author{A.V. Kavokin}
\affiliation{School of Physics and Astronomy, University of Southampton, Southampton, SO17
 1BJ, United Kingdom}
\affiliation{
Spin Optics Laboratory, St-Petersburg State University, 1, Ulianovskaya,
St-Peterbsurg, 198504, Russia}
\author{Z. Hatzopoulos}
\affiliation{Microelectronics Research Group, IESL-FORTH, P.O. Box 1527, 71110
Heraklion, Crete, Greece}
\affiliation{Department of Physics, University of Crete, 71003 Heraklion, Crete, Greece}
\author{P.G. Savvidis}
\affiliation{Microelectronics Research Group, IESL-FORTH, P.O. Box 1527, 71110
Heraklion, Crete, Greece}
\affiliation{Department of Materials Science and Technology, University of Crete, Crete, Greece}
\author{P.G. Lagoudakis}
\affiliation{School of Physics and Astronomy, University of Southampton, Southampton, SO17
 1BJ, United Kingdom}

\begin{abstract}

 We demonstrate experimentally the condensation of exciton-polaritons through optical trapping.
 The non-resonant pump profile is shaped into a ring and projected to a high quality factor microcavity where it forms a 2D repulsive optical potential 
originating from the interactions of polaritons with the excitonic reservoir. Increasing the population of particles in the trap eventually leads to the 
emergence of a confined polariton condensate that is spatially decoupled from the decoherence inducing reservoir, before any build up of coherence on 
the excitation region. In a reference experiment, where the trapping mechanism is switched off by changing the excitation intensity profile, polariton 
condensation takes place for excitation densities more than two times higher and the resulting condensate is subject to a much stronger dephasing and depletion processes.
 \end{abstract}
\maketitle

Strong coupling of cavity photons and quantum-well excitons gives rise to mixed
light-matter bosonic quasiparticles called exciton-polaritons or polaritons
\cite{kavokin_microcavities_2007}. Due to their photonic component, polaritons
are several orders of magnitude lighter than atoms, which makes their
condensation  attainable at higher
temperatures~\cite{kasprzak_bose-einstein_2006,balili_bose-einstein_2007}.  The
manifestations of polariton condensation include polariton lasing
\cite{christopoulos_room-temperature_2007}, long-range spatial coherence
\cite{nardin_dynamics_2009,deng_spatial_2007} and stochastic vector polarisation
\cite{ohadi_spontaneous_2012}.  In an ideal infinite two-dimensional cavity the
polariton gas is expected to undergo the Berezinsky-Kosterlitz-Thouless (BKT)
phase transition~\cite{malpuech_polariton_2003}, while in realistic structures,
polaritons can condense in traps induced by random optical
disorder~\cite{kasprzak_bose-einstein_2006} or mechanically created
potentials~\cite{balili_bose-einstein_2007,balili_actively_2006}.  Polariton
condensation has also been observed in structures of lower
dimensionality~\cite{maragkou_spontaneous_2010,bajoni_polariton_2008,ferrier_interactions_2011,das_polariton_2013}
where the structure itself acts as the trapping potential.  Furthermore, the
manipulation of polariton condensates by optically generated potentials has been
previously shown ~\cite{wertz_spontaneous_2010,tosi_sculpting_2012,gao_2012}. In
these works the condensation process was not assisted by the optical potential
but used to localize an already formed polariton condensate.

Here, we report on the first manifestation of polariton condensation assisted by
an optically generated two dimensional potential. This scheme allows for the
formation of a polariton condensate spatially separated from the excitation
spot. Owing to the efficient trapping in the optical potential we observe a
reduced excitation density threshold as well as higher coherence due to the
decoupling of the condensate from the exciton reservoir. Polariton
condensation prior to the build-up of coherence in the form of photon or
polariton lasing~\cite{kammann_crossover_2012,deng_polariton_2003} at the excitation area on the
sample, decisively resolves the debate on the phase relation between the
excitation laser and polariton condensate.

We used a high quality factor GaAs/AlGaAs microcavity containing four separate
triplets of $\unit[10]{nm}$ GaAs quantum wells and has a vacuum Rabi splitting
of  $\unit[9]{meV}$~\cite{tsotsis_2012}, held at $\sim\unit[6.5]{K}$ in a
cold-finger cryostat and excited non-resonantly at the first reflection minimum
above the cavity stop band with a single-mode continuous wave laser.  The
excitation beam profile was shaped into a ring in real space with the use of two
axicons and was projected to the microcavity through an objective lens
($\mathrm{NA}=0.4$) creating a polariton ring with a mean diameter of
$\sim\unit[20]{\mu m}$ on the sample (Supplementary Information,
\ref{sec:exp_setup}), which is of the order of the polariton mean free path in
planar microcavities and much larger than the exciton diffusion length of the
quantum wells of our
sample~\cite{heller_direct_1996,nagamune_onedimensional_1995}. The excitation
beam intensity was modulated with an acousto-optic modulator at 1\% duty cycle
with a frequency of $\unit[10]{kHz}$ to reduce heating. 
\begin{figure}[h]
\includegraphics[scale=0.85]{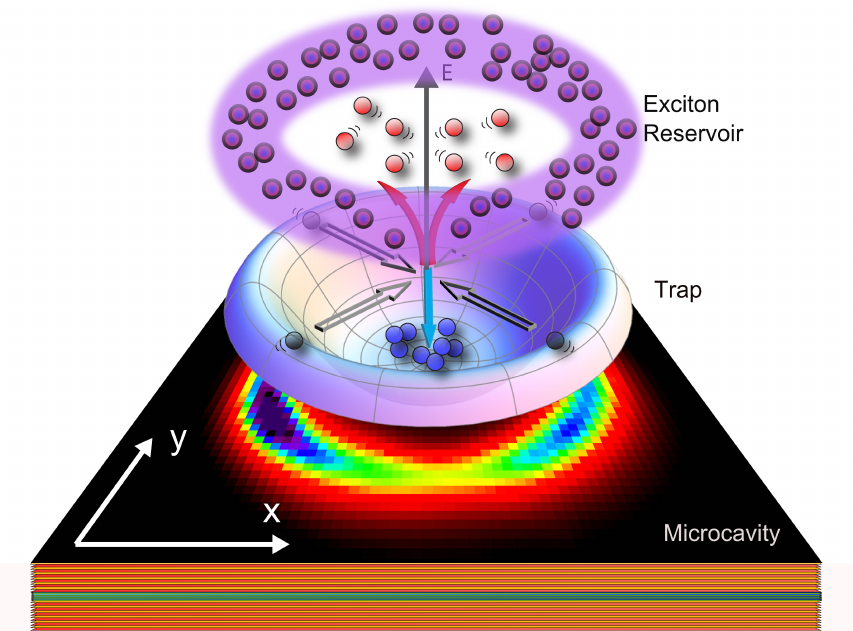}
\centering
\caption{Polariton trap over real space excitation spot, displaying the trapping mechanism. Polaritons scattered to high energy states leave the trap, while those scattered to low energy and momentum remain confined.}
\label{fig:schematic_trap}
\end{figure}

\begin{figure*}
\includegraphics[scale=0.7]{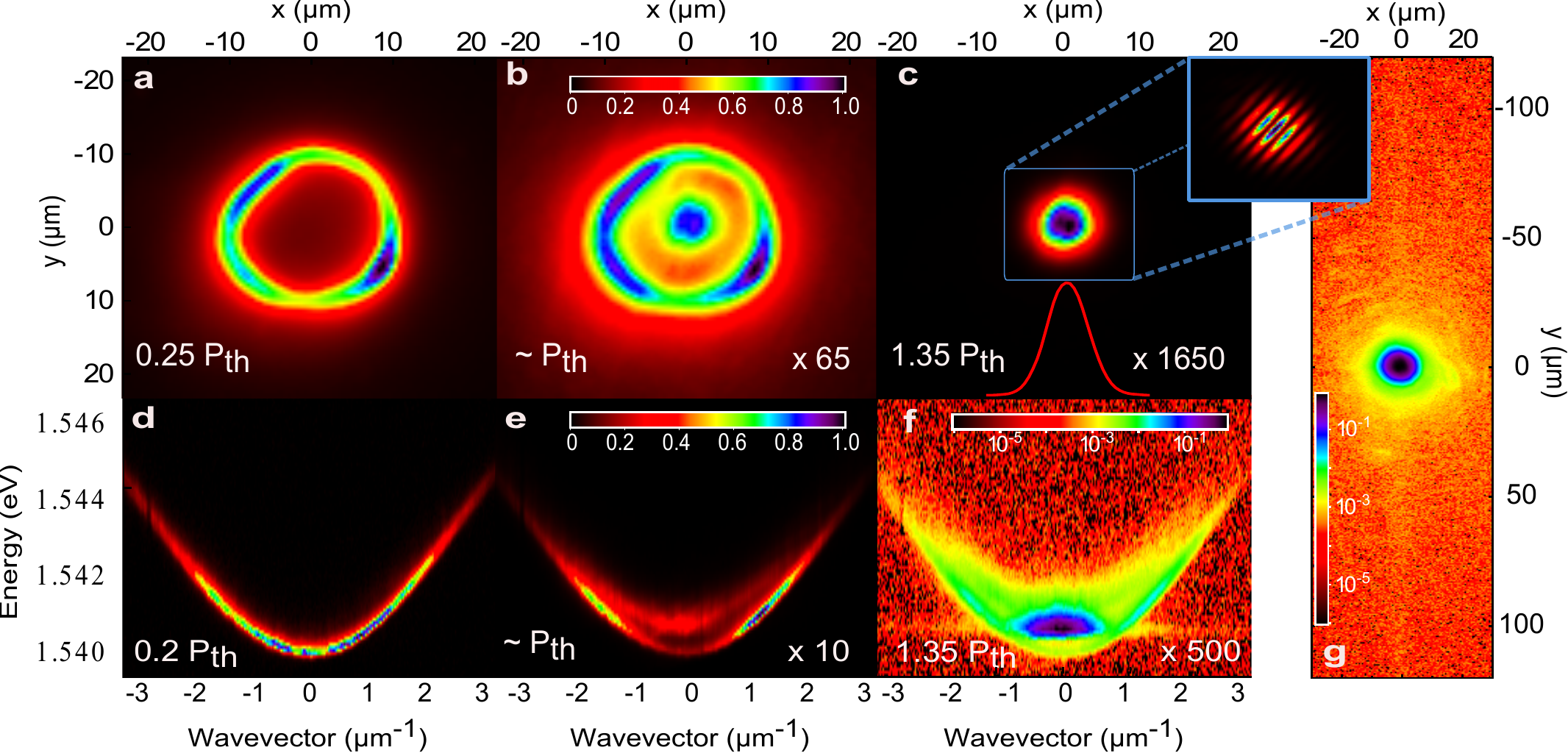}
\centering
\caption{
        Single-mode exciton-polariton condensate in a ring-shaped trap.
        Emission images in real space below (a), at (b) and
        above threshold (c). The inset in (c) shows the
        interference fringes of the condensate. Red line in (c) is the condensate profile
        along the $x$ axis.  Polariton condensation is clearly visible
        in the center of the ring and is separate from the excitation spot
        outlined by the emission in (a). (d), (e), (f) Dispersion images
        below (d), at (e) and above (f) threshold.
        (g), Same as (c) but with a logarithmic colourscale
        showing that the condensate propagation beyond the trap is minimal. 
}
\label{fig:real_k}
\end{figure*}
The nonresonant excitation creates a hot electron-hole plasma, which then forms
excitons. Hot excitons cool down by exciton-phonon
scatterings~\cite{gulia_phonon-assisted_1997}.  When they enter the light cone
they couple strongly to the cavity mode and populate the lower polariton branch
on the ring. Excitons diffuse around the excitation area but due to their large
effective mass they are unable to reach the center of the ring. The repulsion of
polaritons from the ring-shaped exciton reservoir can be described by a
mean-field ring-like trapping potential, which is approximately $\unit[1]{meV}$
deep in the center at the pumping power corresponding to the condensation
threshold. Uncondensed polaritons start from the blueshifted states on the ring
and ballistically expand~\cite{kammann_nonlinear_2012} either towards the center
or outside. Those which propagate to the center eventually collide with each
other (see Fig~\ref{fig:schematic_trap}). The energy of the ensemble of
polaritons is conserved by these scattering events, so that the kinetic energies
of approximately half of the polaritons are reduced, while the other half have
their kinetic energies increased. As a result, a fraction of the polariton gas
is no more capable to escape from the trap due to the lack of kinetic energy,
while the rest can easily fly away over the barriers.  Further scatterings of
the trapped polaritons lead to the increase of the kinetic energy of some of
them so that they become able to leave the trap. By increasing the excitation
power, the polariton population inside the trap builds up and a condensate forms
at the center of the ring that is quickly enhanced due to final state polariton
stimulated scatterings~\cite{savvidis_angle-resonant_2000}.

Polariton emission in real space for powers greatly below threshold outlines the
pump profile (Fig.~\ref{fig:real_k}a).  At the onset of condensation,
photoluminescence(PL) from the center of the trap is of the same intensity as
emission from the ring  (Fig.~\ref{fig:real_k}b). Above threshold
(Fig.~\ref{fig:real_k}c,g) a Gaussian shaped single-mode condensate, with full
width at half maximum (FWHM) of $\unit[5.46]{\mu m}$ and standard deviation
$\sigma_x=\unit[2.32]{\mu m}$, is formed and effectively confined inside the
ring (images of the complete power dependence have been compiled in a video that
can be found in the supplementary information). Michelson interferometry images
(inset in Fig.~\ref{fig:real_k}c) confirm the buildup of coherence in the
condensate (Supplementary Information,~\ref{sec:interfer}). 

The dispersion of polaritons for the entire surface of the ring and for
different pumping powers  can be seen at Fig.~\ref{fig:real_k}d-f. Below
threshold we observe a normal parabolic lower polariton branch. As the polariton
density in the centre of the ring is increased close to threshold, we observe a
blueshifted dispersion from the polaritons in the trap, coexisting with the
parabolic dispersion of untrapped polaritons as it will become evident further
on in this Communication from spatially resolved dispersion imaging. The two
lobes of the outer dispersion in Fig.~\ref{fig:real_k}e correspond to high
momentum polaritons escaping from the centre of the ring. By further increasing
the excitation power a condensate appears in the ground state of the blueshifted
dispersion with zero in-plane momentum and standard deviation $
\sigma_{k_x}=\unit[0.24]{\mu m^{-1}}$, shown in logarithmic scale in
Fig.~\ref{fig:real_k}f. The macroscopically occupied ground state is very close
to the Heisenberg limit, having $\sigma_x \sigma_{k_x}=0.56$ lower than
previously reported values~\cite{roumpos_gain-induced_2010}. This confirms that
phase fluctuations in the condensate are strongly reduced.

\begin{figure}
\includegraphics[scale=0.75]{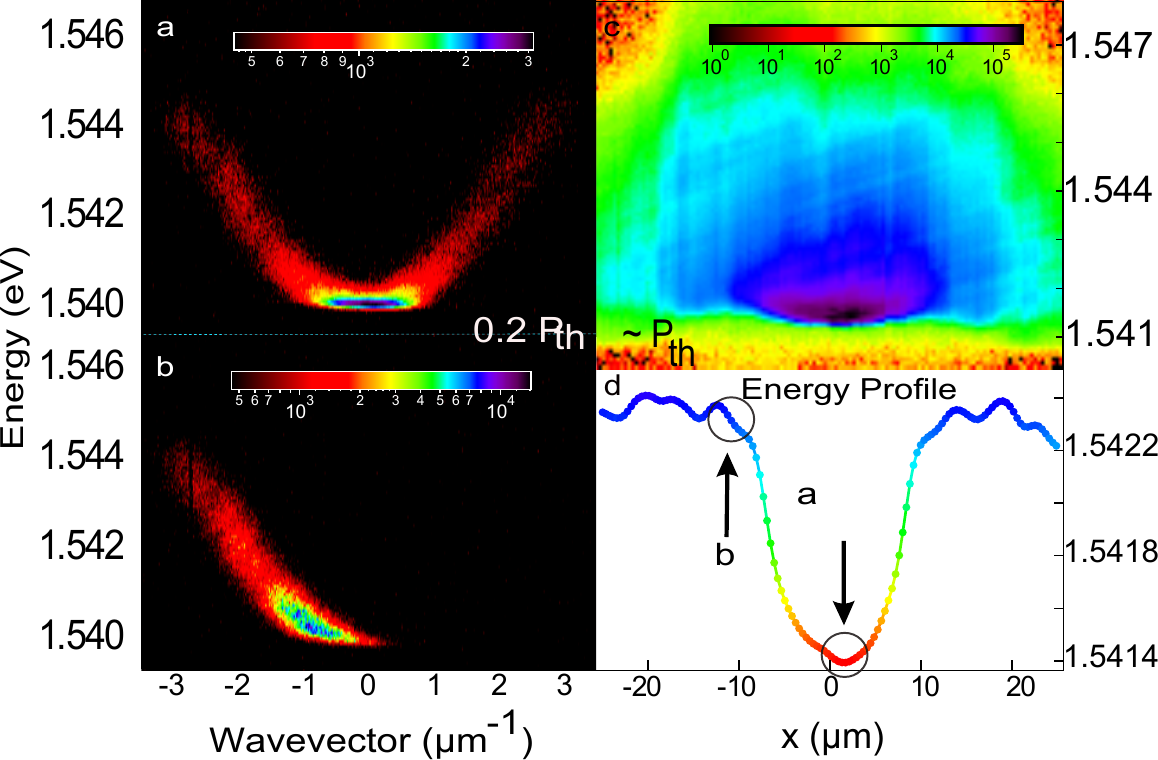}
\centering
\caption{
    Trapping potential and spatially-resolved dispersions. (a),
    (b), Dispersion images below threshold ($P= 0.2 P_{th}$) from a $ \unit[5]{\mu
    m}$ diameter spatially filtered region from the center of the ring (a)
    and from the rim (b). Energy profile of the central slice of the ring
    at threshold (c) and the extracted maxima of intensity
    along the $x$ axis for visualization of the trap profile (d). Points a and b correspond to (a) and (b) respectively.}
\label{fig:sf_dispersions}
\end{figure}

Spatially-resolved dispersion images reveal that untrapped polaritons positioned
on the rim of the ring have high energies and large wavevectors
(Fig.~\ref{fig:sf_dispersions}b), while those in the center of the trap
primarily populate the lower states even at pump powers much below threshold
(Fig.~\ref{fig:sf_dispersions}a) (see also supplementary information
\ref{sec:op_trap}). The dispersion of the polaritons on the edge of the
ring does not change greatly with increasing power(for the power range that we
examined), while the dispersion images at the center of the trap demonstrate
condensation at $k=0$ above threshold.  The profile of the trap can be
visualized by energy resolving the central slice ($ <\unit[0.2]{\mu m}$) of the
excitation ring (Fig.~\ref{fig:sf_dispersions}c). By extracting the energy that
corresponds to the maximum intensity along each point of the $x$ axis of
Fig.~\ref{fig:sf_dispersions}c, the trap potential can be assembled
(Fig.~\ref{fig:sf_dispersions}d).  The trap depth at threshold is $\sim
\unit[1]{meV}$. The two circles in Fig.~\ref{fig:sf_dispersions}d annotate the
points where the spatially filtered dispersions were acquired.  

\begin{figure}[h]
\includegraphics[scale=0.67]{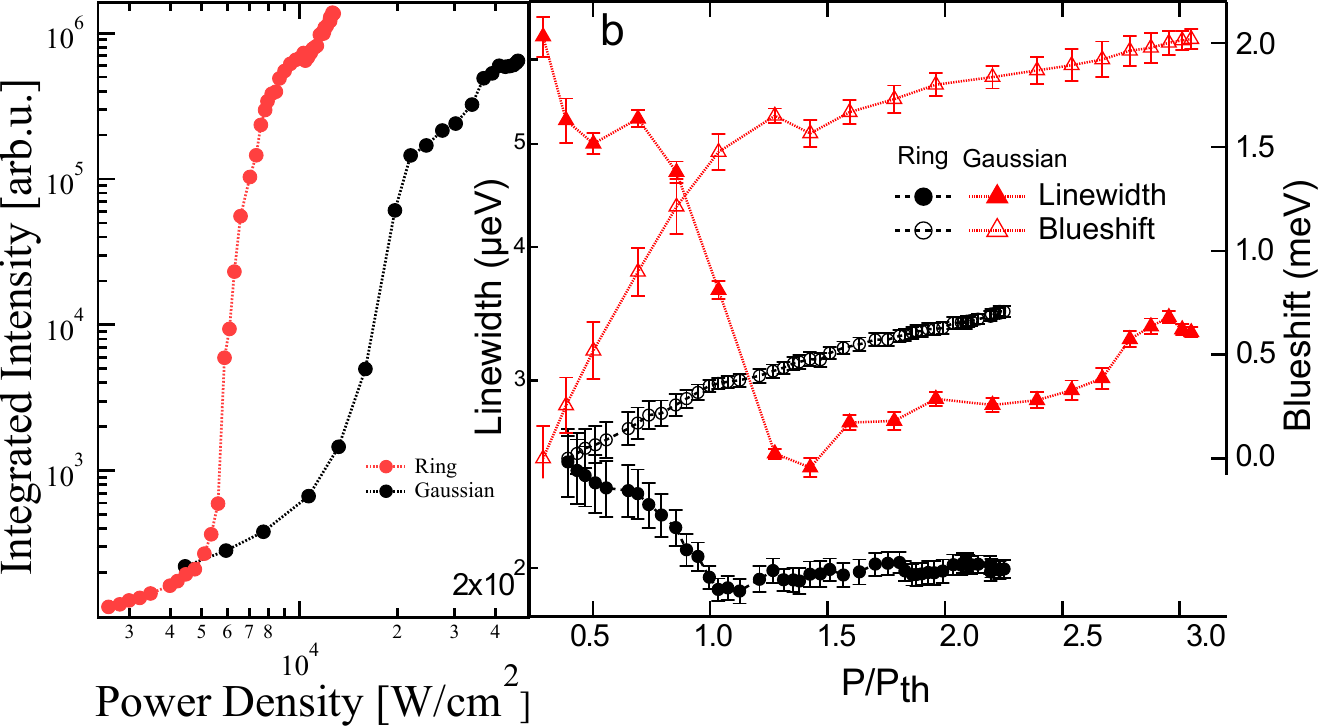}
\centering
\caption{
    Normalized intensity, linewidth and blueshif. (a),
    Normalized intensity with increasing power density for polaritons in the
    center (open black dots) of the ring excitation, and for a Gaussian excitation (open red circles) measured
    at $k$=$0\pm\unit[0.05]{\mu m^{-1}}$ of the spatially filtered dispersion.
    (b), Linewidth (in log scale) and blueshift (right axis) vs. excitation power for polaritons at
    the center of the ring (black circles, open black circles respectively) and
    a Gaussian excitation (red triangles, open red triangles respectively).
    Lines are guides for the eyes.
}
\label{fig:linewidth_blueshift}
\end{figure}

Full spatial separation of the condensate from the pump induced excitonic
reservoir has important implications on the spectral and dynamic properties of
polaritons~\cite{vishnevsky_multistability_2012} even below threshold. Due to
the efficient stimulated scattering process we observe lower power densities for
condensation. In a reference experiment we have excited the same sample (at the
same detuning and temperature) with a normal Gaussian beam of spot size of $\sim
\unit[5]{\mu m}$.  Fig.~\ref{fig:linewidth_blueshift}a shows the integrated PL
peak intensity of $k=0$ for different excitation powers. The threshold power
density is more than two times higher in the case of Gaussian excitation. 

A clear advantage of separation of the condensate from the feeding reservoir is
in the strong reduction of the depletion processes caused by
condensate-reservoir interactions~\cite{love_intrinsic_2008}.
Fig.~\ref{fig:linewidth_blueshift}b shows the linewidth and blueshift of the
condensate for the ring and Gaussian excitation cases. Due to the absence of
decoherence mechanisms by the background reservoir in the case of
ring-excitation, the linewidth is narrower and it increases much slower than in
the Gaussian excitation. In the case of the ring excitation the dephasing of the
condensate due the interaction with the exciton reservoir  is strongly
suppressed~\cite{porras_linewidth_2003}.  The blueshift of the condensate
increases linearly with the pumping intensity, in the case of ring-like
excitation, showing the linear increase of the mean number of condensed
polaritons.  In the case of Gaussian excitation, the blue shift is strongly
affected by the reservoir: it is twice as large as in the ring-excitation case
and slowly saturates above threshold, indicating that exciton saturation has
been reached~\cite{holden_exciton_1997}.

In conclusion, we have demonstrated the condensation of a polariton bosonic gas
in a two-dimensional optical trap. This configuration allows for the formation
of a polariton condensate spatially separated from the excitation area
minimizing dephasing and depletion processes associated with the excitonic
reservoir. This highly efficient excitation technique of exciton-polaritons
results in the spontaneous formation of a polariton BEC spatially separated from
the excitation laser and at more than two times lower excitation densities
compared to previous experimental configurations. In the case of a polariton BEC
formed through optical trapping the linewidth reduces and clamps at threshold
clearly evidencing that temporal coherence is not affected by increasing the
occupation number of the condensate. Disassociation of the condensate from the
excitation beam, conclusively settles the debate on the inheritance of coherence
of the polariton condensate from the excitation laser. 

We acknowledge funding from Marie Curie ITNs Spinoptronics, Clermont IV and
EPSRC through Contract No. EP/F026455/1. P.G.S. acknowledges funding from the
EU Social Fund and Greek National Resources (EPEAEK II, HRAKLEITOS II). 


\begin{thebibliography}{29}%
\makeatletter
\providecommand \@ifxundefined [1]{%
 \@ifx{#1\undefined}
}%
\providecommand \@ifnum [1]{%
 \ifnum #1\expandafter \@firstoftwo
 \else \expandafter \@secondoftwo
 \fi
}%
\providecommand \@ifx [1]{%
 \ifx #1\expandafter \@firstoftwo
 \else \expandafter \@secondoftwo
 \fi
}%
\providecommand \natexlab [1]{#1}%
\providecommand \enquote  [1]{``#1''}%
\providecommand \bibnamefont  [1]{#1}%
\providecommand \bibfnamefont [1]{#1}%
\providecommand \citenamefont [1]{#1}%
\providecommand \href@noop [0]{\@secondoftwo}%
\providecommand \href [0]{\begingroup \@sanitize@url \@href}%
\providecommand \@href[1]{\@@startlink{#1}\@@href}%
\providecommand \@@href[1]{\endgroup#1\@@endlink}%
\providecommand \@sanitize@url [0]{\catcode `\\12\catcode `\$12\catcode
  `\&12\catcode `\#12\catcode `\^12\catcode `\_12\catcode `\%12\relax}%
\providecommand \@@startlink[1]{}%
\providecommand \@@endlink[0]{}%
\providecommand \url  [0]{\begingroup\@sanitize@url \@url }%
\providecommand \@url [1]{\endgroup\@href {#1}{\urlprefix }}%
\providecommand \urlprefix  [0]{URL }%
\providecommand \Eprint [0]{\href }%
\providecommand \doibase [0]{http://dx.doi.org/}%
\providecommand \selectlanguage [0]{\@gobble}%
\providecommand \bibinfo  [0]{\@secondoftwo}%
\providecommand \bibfield  [0]{\@secondoftwo}%
\providecommand \translation [1]{[#1]}%
\providecommand \BibitemOpen [0]{}%
\providecommand \bibitemStop [0]{}%
\providecommand \bibitemNoStop [0]{.\EOS\space}%
\providecommand \EOS [0]{\spacefactor3000\relax}%
\providecommand \BibitemShut  [1]{\csname bibitem#1\endcsname}%
\let\auto@bib@innerbib\@empty
\bibitem [{\citenamefont {Kavokin}\ \emph {et~al.}(2007)\citenamefont
  {Kavokin}, \citenamefont {Baumberg}, \citenamefont {Malpuech},\ and\
  \citenamefont {Laussy}}]{kavokin_microcavities_2007}%
  \BibitemOpen
  \bibfield  {author} {\bibinfo {author} {\bibfnamefont {A.~V.}\ \bibnamefont
  {Kavokin}}, \bibinfo {author} {\bibfnamefont {J.}~\bibnamefont {Baumberg}},
  \bibinfo {author} {\bibfnamefont {G.}~\bibnamefont {Malpuech}}, \ and\
  \bibinfo {author} {\bibfnamefont {F.~P.}\ \bibnamefont {Laussy}},\
  }\href@noop {} {\emph {\bibinfo {title} {Microcavities}}}\ (\bibinfo
  {publisher} {Oxford Univ. Press},\ \bibinfo {address} {Oxford},\ \bibinfo
  {year} {2007})\BibitemShut {NoStop}%
\bibitem [{\citenamefont {Kasprzak}\ \emph {et~al.}(2006)\citenamefont
  {Kasprzak}, \citenamefont {Richard}, \citenamefont {Kundermann},
  \citenamefont {Baas}, \citenamefont {Jeambrun}, \citenamefont {Keeling},
  \citenamefont {Marchetti}, \citenamefont {Szymańska}, \citenamefont
  {André}, \citenamefont {Staehli}, \citenamefont {Savona}, \citenamefont
  {Littlewood}, \citenamefont {Deveaud},\ and\ \citenamefont
  {Dang}}]{kasprzak_bose-einstein_2006}%
  \BibitemOpen
  \bibfield  {author} {\bibinfo {author} {\bibfnamefont {J.}~\bibnamefont
  {Kasprzak}}, \bibinfo {author} {\bibfnamefont {M.}~\bibnamefont {Richard}},
  \bibinfo {author} {\bibfnamefont {S.}~\bibnamefont {Kundermann}}, \bibinfo
  {author} {\bibfnamefont {A.}~\bibnamefont {Baas}}, \bibinfo {author}
  {\bibfnamefont {P.}~\bibnamefont {Jeambrun}}, \bibinfo {author}
  {\bibfnamefont {J.~M.~J.}\ \bibnamefont {Keeling}}, \bibinfo {author}
  {\bibfnamefont {F.~M.}\ \bibnamefont {Marchetti}}, \bibinfo {author}
  {\bibfnamefont {M.~H.}\ \bibnamefont {Szymańska}}, \bibinfo {author}
  {\bibfnamefont {R.}~\bibnamefont {André}}, \bibinfo {author} {\bibfnamefont
  {J.~L.}\ \bibnamefont {Staehli}}, \bibinfo {author} {\bibfnamefont
  {V.}~\bibnamefont {Savona}}, \bibinfo {author} {\bibfnamefont {P.~B.}\
  \bibnamefont {Littlewood}}, \bibinfo {author} {\bibfnamefont
  {B.}~\bibnamefont {Deveaud}}, \ and\ \bibinfo {author} {\bibfnamefont
  {L.~S.}\ \bibnamefont {Dang}},\ }\href {\doibase 10.1038/nature05131}
  {\bibfield  {journal} {\bibinfo  {journal} {Nature}\ }\textbf {\bibinfo
  {volume} {443}},\ \bibinfo {pages} {409–14} (\bibinfo {year}
  {2006})}\BibitemShut {NoStop}%
\bibitem [{\citenamefont {Balili}\ \emph {et~al.}(2007)\citenamefont {Balili},
  \citenamefont {Hartwell}, \citenamefont {Snoke}, \citenamefont {Pfeiffer},\
  and\ \citenamefont {West}}]{balili_bose-einstein_2007}%
  \BibitemOpen
  \bibfield  {author} {\bibinfo {author} {\bibfnamefont {R.}~\bibnamefont
  {Balili}}, \bibinfo {author} {\bibfnamefont {V.}~\bibnamefont {Hartwell}},
  \bibinfo {author} {\bibfnamefont {D.}~\bibnamefont {Snoke}}, \bibinfo
  {author} {\bibfnamefont {L.}~\bibnamefont {Pfeiffer}}, \ and\ \bibinfo
  {author} {\bibfnamefont {K.}~\bibnamefont {West}},\ }\href {\doibase
  10.1126/science.1140990} {\bibfield  {journal} {\bibinfo  {journal}
  {Science}\ }\textbf {\bibinfo {volume} {316}},\ \bibinfo {pages} {1007 }
  (\bibinfo {year} {2007})}\BibitemShut {NoStop}%
\bibitem [{\citenamefont {Christopoulos}\ \emph {et~al.}(2007)\citenamefont
  {Christopoulos}, \citenamefont {von Högersthal}, \citenamefont {Grundy},
  \citenamefont {Lagoudakis}, \citenamefont {Kavokin}, \citenamefont
  {Baumberg}, \citenamefont {Christmann}, \citenamefont {Butté}, \citenamefont
  {Feltin}, \citenamefont {Carlin},\ and\ \citenamefont
  {Grandjean}}]{christopoulos_room-temperature_2007}%
  \BibitemOpen
  \bibfield  {author} {\bibinfo {author} {\bibfnamefont {S.}~\bibnamefont
  {Christopoulos}}, \bibinfo {author} {\bibfnamefont {G.~B.~H.}\ \bibnamefont
  {von Högersthal}}, \bibinfo {author} {\bibfnamefont {A.~J.~D.}\ \bibnamefont
  {Grundy}}, \bibinfo {author} {\bibfnamefont {P.~G.}\ \bibnamefont
  {Lagoudakis}}, \bibinfo {author} {\bibfnamefont {A.~V.}\ \bibnamefont
  {Kavokin}}, \bibinfo {author} {\bibfnamefont {J.~J.}\ \bibnamefont
  {Baumberg}}, \bibinfo {author} {\bibfnamefont {G.}~\bibnamefont
  {Christmann}}, \bibinfo {author} {\bibfnamefont {R.}~\bibnamefont {Butté}},
  \bibinfo {author} {\bibfnamefont {E.}~\bibnamefont {Feltin}}, \bibinfo
  {author} {\bibfnamefont {J.-F.}\ \bibnamefont {Carlin}}, \ and\ \bibinfo
  {author} {\bibfnamefont {N.}~\bibnamefont {Grandjean}},\ }\href {\doibase
  10.1103/PhysRevLett.98.126405} {\bibfield  {journal} {\bibinfo  {journal}
  {Physical Review Letters}\ }\textbf {\bibinfo {volume} {98}},\ \bibinfo
  {pages} {126405} (\bibinfo {year} {2007})}\BibitemShut {NoStop}%
\bibitem [{\citenamefont {Nardin}\ \emph {et~al.}(2009)\citenamefont {Nardin},
  \citenamefont {Lagoudakis}, \citenamefont {Wouters}, \citenamefont {Richard},
  \citenamefont {Baas}, \citenamefont {André}, \citenamefont {Dang},
  \citenamefont {Pietka},\ and\ \citenamefont
  {Deveaud-Plédran}}]{nardin_dynamics_2009}%
  \BibitemOpen
  \bibfield  {author} {\bibinfo {author} {\bibfnamefont {G.}~\bibnamefont
  {Nardin}}, \bibinfo {author} {\bibfnamefont {K.~G.}\ \bibnamefont
  {Lagoudakis}}, \bibinfo {author} {\bibfnamefont {M.}~\bibnamefont {Wouters}},
  \bibinfo {author} {\bibfnamefont {M.}~\bibnamefont {Richard}}, \bibinfo
  {author} {\bibfnamefont {A.}~\bibnamefont {Baas}}, \bibinfo {author}
  {\bibfnamefont {R.}~\bibnamefont {André}}, \bibinfo {author} {\bibfnamefont
  {L.~S.}\ \bibnamefont {Dang}}, \bibinfo {author} {\bibfnamefont
  {B.}~\bibnamefont {Pietka}}, \ and\ \bibinfo {author} {\bibfnamefont
  {B.}~\bibnamefont {Deveaud-Plédran}},\ }\href {\doibase
  10.1103/PhysRevLett.103.256402} {\bibfield  {journal} {\bibinfo  {journal}
  {Physical Review Letters}\ }\textbf {\bibinfo {volume} {103}},\ \bibinfo
  {pages} {256402} (\bibinfo {year} {2009})}\BibitemShut {NoStop}%
\bibitem [{\citenamefont {Deng}\ \emph {et~al.}(2007)\citenamefont {Deng},
  \citenamefont {Solomon}, \citenamefont {Hey}, \citenamefont {Ploog},\ and\
  \citenamefont {Yamamoto}}]{deng_spatial_2007}%
  \BibitemOpen
  \bibfield  {author} {\bibinfo {author} {\bibfnamefont {H.}~\bibnamefont
  {Deng}}, \bibinfo {author} {\bibfnamefont {G.~S.}\ \bibnamefont {Solomon}},
  \bibinfo {author} {\bibfnamefont {R.}~\bibnamefont {Hey}}, \bibinfo {author}
  {\bibfnamefont {K.~H.}\ \bibnamefont {Ploog}}, \ and\ \bibinfo {author}
  {\bibfnamefont {Y.}~\bibnamefont {Yamamoto}},\ }\href {\doibase
  10.1103/PhysRevLett.99.126403} {\bibfield  {journal} {\bibinfo  {journal}
  {Physical Review Letters}\ }\textbf {\bibinfo {volume} {99}},\ \bibinfo
  {pages} {126403} (\bibinfo {year} {2007})}\BibitemShut {NoStop}%
\bibitem [{\citenamefont {Ohadi}\ \emph {et~al.}(2012)\citenamefont {Ohadi},
  \citenamefont {Kammann}, \citenamefont {Liew}, \citenamefont {Lagoudakis},
  \citenamefont {Kavokin},\ and\ \citenamefont
  {Lagoudakis}}]{ohadi_spontaneous_2012}%
  \BibitemOpen
  \bibfield  {author} {\bibinfo {author} {\bibfnamefont {H.}~\bibnamefont
  {Ohadi}}, \bibinfo {author} {\bibfnamefont {E.}~\bibnamefont {Kammann}},
  \bibinfo {author} {\bibfnamefont {T.~C.~H.}\ \bibnamefont {Liew}}, \bibinfo
  {author} {\bibfnamefont {K.~G.}\ \bibnamefont {Lagoudakis}}, \bibinfo
  {author} {\bibfnamefont {A.~V.}\ \bibnamefont {Kavokin}}, \ and\ \bibinfo
  {author} {\bibfnamefont {P.~G.}\ \bibnamefont {Lagoudakis}},\ }\href
  {\doibase 10.1103/PhysRevLett.109.016404} {\bibfield  {journal} {\bibinfo
  {journal} {Physical Review Letters}\ }\textbf {\bibinfo {volume} {109}},\
  \bibinfo {pages} {016404} (\bibinfo {year} {2012})}\BibitemShut {NoStop}%
\bibitem [{\citenamefont {Malpuech}\ \emph {et~al.}(2003)\citenamefont
  {Malpuech}, \citenamefont {Rubo}, \citenamefont {Laussy}, \citenamefont
  {Bigenwald},\ and\ \citenamefont {Kavokin}}]{malpuech_polariton_2003}%
  \BibitemOpen
  \bibfield  {author} {\bibinfo {author} {\bibfnamefont {G.}~\bibnamefont
  {Malpuech}}, \bibinfo {author} {\bibfnamefont {Y.~G.}\ \bibnamefont {Rubo}},
  \bibinfo {author} {\bibfnamefont {F.~P.}\ \bibnamefont {Laussy}}, \bibinfo
  {author} {\bibfnamefont {P.}~\bibnamefont {Bigenwald}}, \ and\ \bibinfo
  {author} {\bibfnamefont {A.~V.}\ \bibnamefont {Kavokin}},\ }\href {\doibase
  10.1088/0268-1242/18/10/314} {\bibfield  {journal} {\bibinfo  {journal}
  {Semiconductor Science and Technology}\ }\textbf {\bibinfo {volume} {18}},\
  \bibinfo {pages} {S395} (\bibinfo {year} {2003})}\BibitemShut {NoStop}%
\bibitem [{\citenamefont {Balili}\ \emph {et~al.}(2006)\citenamefont {Balili},
  \citenamefont {Snoke}, \citenamefont {Pfeiffer},\ and\ \citenamefont
  {West}}]{balili_actively_2006}%
  \BibitemOpen
  \bibfield  {author} {\bibinfo {author} {\bibfnamefont {R.~B.}\ \bibnamefont
  {Balili}}, \bibinfo {author} {\bibfnamefont {D.~W.}\ \bibnamefont {Snoke}},
  \bibinfo {author} {\bibfnamefont {L.}~\bibnamefont {Pfeiffer}}, \ and\
  \bibinfo {author} {\bibfnamefont {K.}~\bibnamefont {West}},\ }\href {\doibase
  doi:10.1063/1.2164431} {\bibfield  {journal} {\bibinfo  {journal} {Applied
  Physics Letters}\ }\textbf {\bibinfo {volume} {88}},\ \bibinfo {pages}
  {031110} (\bibinfo {year} {2006})}\BibitemShut {NoStop}%
\bibitem [{\citenamefont {Maragkou}\ \emph {et~al.}(2010)\citenamefont
  {Maragkou}, \citenamefont {Grundy}, \citenamefont {Wertz}, \citenamefont
  {Lemaître}, \citenamefont {Sagnes}, \citenamefont {Senellart}, \citenamefont
  {Bloch},\ and\ \citenamefont {Lagoudakis}}]{maragkou_spontaneous_2010}%
  \BibitemOpen
  \bibfield  {author} {\bibinfo {author} {\bibfnamefont {M.}~\bibnamefont
  {Maragkou}}, \bibinfo {author} {\bibfnamefont {A.~J.~D.}\ \bibnamefont
  {Grundy}}, \bibinfo {author} {\bibfnamefont {E.}~\bibnamefont {Wertz}},
  \bibinfo {author} {\bibfnamefont {A.}~\bibnamefont {Lemaître}}, \bibinfo
  {author} {\bibfnamefont {I.}~\bibnamefont {Sagnes}}, \bibinfo {author}
  {\bibfnamefont {P.}~\bibnamefont {Senellart}}, \bibinfo {author}
  {\bibfnamefont {J.}~\bibnamefont {Bloch}}, \ and\ \bibinfo {author}
  {\bibfnamefont {P.~G.}\ \bibnamefont {Lagoudakis}},\ }\href {\doibase
  10.1103/PhysRevB.81.081307} {\bibfield  {journal} {\bibinfo  {journal}
  {Physical Review B}\ }\textbf {\bibinfo {volume} {81}},\ \bibinfo {pages}
  {081307} (\bibinfo {year} {2010})}\BibitemShut {NoStop}%
\bibitem [{\citenamefont {Bajoni}\ \emph {et~al.}(2008)\citenamefont {Bajoni},
  \citenamefont {Senellart}, \citenamefont {Wertz}, \citenamefont {Sagnes},
  \citenamefont {Miard}, \citenamefont {Lemaître},\ and\ \citenamefont
  {Bloch}}]{bajoni_polariton_2008}%
  \BibitemOpen
  \bibfield  {author} {\bibinfo {author} {\bibfnamefont {D.}~\bibnamefont
  {Bajoni}}, \bibinfo {author} {\bibfnamefont {P.}~\bibnamefont {Senellart}},
  \bibinfo {author} {\bibfnamefont {E.}~\bibnamefont {Wertz}}, \bibinfo
  {author} {\bibfnamefont {I.}~\bibnamefont {Sagnes}}, \bibinfo {author}
  {\bibfnamefont {A.}~\bibnamefont {Miard}}, \bibinfo {author} {\bibfnamefont
  {A.}~\bibnamefont {Lemaître}}, \ and\ \bibinfo {author} {\bibfnamefont
  {J.}~\bibnamefont {Bloch}},\ }\href {\doibase 10.1103/PhysRevLett.100.047401}
  {\bibfield  {journal} {\bibinfo  {journal} {Physical Review Letters}\
  }\textbf {\bibinfo {volume} {100}},\ \bibinfo {pages} {047401} (\bibinfo
  {year} {2008})}\BibitemShut {NoStop}%
\bibitem [{\citenamefont {Ferrier}\ \emph {et~al.}(2011)\citenamefont
  {Ferrier}, \citenamefont {Wertz}, \citenamefont {Johne}, \citenamefont
  {Solnyshkov}, \citenamefont {Senellart}, \citenamefont {Sagnes},
  \citenamefont {Lemaître}, \citenamefont {Malpuech},\ and\ \citenamefont
  {Bloch}}]{ferrier_interactions_2011}%
  \BibitemOpen
  \bibfield  {author} {\bibinfo {author} {\bibfnamefont {L.}~\bibnamefont
  {Ferrier}}, \bibinfo {author} {\bibfnamefont {E.}~\bibnamefont {Wertz}},
  \bibinfo {author} {\bibfnamefont {R.}~\bibnamefont {Johne}}, \bibinfo
  {author} {\bibfnamefont {D.~D.}\ \bibnamefont {Solnyshkov}}, \bibinfo
  {author} {\bibfnamefont {P.}~\bibnamefont {Senellart}}, \bibinfo {author}
  {\bibfnamefont {I.}~\bibnamefont {Sagnes}}, \bibinfo {author} {\bibfnamefont
  {A.}~\bibnamefont {Lemaître}}, \bibinfo {author} {\bibfnamefont
  {G.}~\bibnamefont {Malpuech}}, \ and\ \bibinfo {author} {\bibfnamefont
  {J.}~\bibnamefont {Bloch}},\ }\href {\doibase 10.1103/PhysRevLett.106.126401}
  {\bibfield  {journal} {\bibinfo  {journal} {Physical Review Letters}\
  }\textbf {\bibinfo {volume} {106}},\ \bibinfo {pages} {126401} (\bibinfo
  {year} {2011})}\BibitemShut {NoStop}%
\bibitem [{\citenamefont {Das}\ \emph {et~al.}(2013)\citenamefont {Das},
  \citenamefont {Bhattacharya}, \citenamefont {Heo}, \citenamefont {Banerjee},\
  and\ \citenamefont {Guo}}]{das_polariton_2013}%
  \BibitemOpen
  \bibfield  {author} {\bibinfo {author} {\bibfnamefont {A.}~\bibnamefont
  {Das}}, \bibinfo {author} {\bibfnamefont {P.}~\bibnamefont {Bhattacharya}},
  \bibinfo {author} {\bibfnamefont {J.}~\bibnamefont {Heo}}, \bibinfo {author}
  {\bibfnamefont {A.}~\bibnamefont {Banerjee}}, \ and\ \bibinfo {author}
  {\bibfnamefont {W.}~\bibnamefont {Guo}},\ }\href {\doibase
  10.1073/pnas.1210842110} {\bibfield  {journal} {\bibinfo  {journal}
  {Proceedings of the National Academy of Sciences}\ }\textbf {\bibinfo
  {volume} {110}},\ \bibinfo {pages} {2735} (\bibinfo {year} {2013})},\
  \bibinfo {note} {{PMID:} 23382183}\BibitemShut {NoStop}%
\bibitem [{\citenamefont {Wertz}\ \emph {et~al.}(2010)\citenamefont {Wertz},
  \citenamefont {Ferrier}, \citenamefont {Solnyshkov}, \citenamefont {Johne},
  \citenamefont {Sanvitto}, \citenamefont {Lemaître}, \citenamefont {Sagnes},
  \citenamefont {Grousson}, \citenamefont {Kavokin}, \citenamefont {Senellart},
  \citenamefont {Malpuech},\ and\ \citenamefont
  {Bloch}}]{wertz_spontaneous_2010}%
  \BibitemOpen
  \bibfield  {author} {\bibinfo {author} {\bibfnamefont {E.}~\bibnamefont
  {Wertz}}, \bibinfo {author} {\bibfnamefont {L.}~\bibnamefont {Ferrier}},
  \bibinfo {author} {\bibfnamefont {D.~D.}\ \bibnamefont {Solnyshkov}},
  \bibinfo {author} {\bibfnamefont {R.}~\bibnamefont {Johne}}, \bibinfo
  {author} {\bibfnamefont {D.}~\bibnamefont {Sanvitto}}, \bibinfo {author}
  {\bibfnamefont {A.}~\bibnamefont {Lemaître}}, \bibinfo {author}
  {\bibfnamefont {I.}~\bibnamefont {Sagnes}}, \bibinfo {author} {\bibfnamefont
  {R.}~\bibnamefont {Grousson}}, \bibinfo {author} {\bibfnamefont {A.~V.}\
  \bibnamefont {Kavokin}}, \bibinfo {author} {\bibfnamefont {P.}~\bibnamefont
  {Senellart}}, \bibinfo {author} {\bibfnamefont {G.}~\bibnamefont {Malpuech}},
  \ and\ \bibinfo {author} {\bibfnamefont {J.}~\bibnamefont {Bloch}},\ }\href
  {\doibase 10.1038/nphys1750} {\bibfield  {journal} {\bibinfo  {journal}
  {Nature Physics}\ }\textbf {\bibinfo {volume} {6}},\ \bibinfo {pages} {860}
  (\bibinfo {year} {2010})}\BibitemShut {NoStop}%
\bibitem [{\citenamefont {Tosi}\ \emph {et~al.}(2012)\citenamefont {Tosi},
  \citenamefont {Christmann}, \citenamefont {Berloff}, \citenamefont {Tsotsis},
  \citenamefont {Gao}, \citenamefont {Hatzopoulos}, \citenamefont {Savvidis},\
  and\ \citenamefont {Baumberg}}]{tosi_sculpting_2012}%
  \BibitemOpen
  \bibfield  {author} {\bibinfo {author} {\bibfnamefont {G.}~\bibnamefont
  {Tosi}}, \bibinfo {author} {\bibfnamefont {G.}~\bibnamefont {Christmann}},
  \bibinfo {author} {\bibfnamefont {N.~G.}\ \bibnamefont {Berloff}}, \bibinfo
  {author} {\bibfnamefont {P.}~\bibnamefont {Tsotsis}}, \bibinfo {author}
  {\bibfnamefont {T.}~\bibnamefont {Gao}}, \bibinfo {author} {\bibfnamefont
  {Z.}~\bibnamefont {Hatzopoulos}}, \bibinfo {author} {\bibfnamefont {P.~G.}\
  \bibnamefont {Savvidis}}, \ and\ \bibinfo {author} {\bibfnamefont {J.~J.}\
  \bibnamefont {Baumberg}},\ }\href {\doibase 10.1038/nphys2182} {\bibfield
  {journal} {\bibinfo  {journal} {Nat Phys}\ }\textbf {\bibinfo {volume}
  {advance online publication}} (\bibinfo {year} {2012}),\
  10.1038/nphys2182}\BibitemShut {NoStop}%
\bibitem [{\citenamefont {Gao}\ \emph {et~al.}(2012)\citenamefont {Gao},
  \citenamefont {Eldridge}, \citenamefont {Liew}, \citenamefont {Tsintzos},
  \citenamefont {Stavrinidis}, \citenamefont {Deligeorgis}, \citenamefont
  {Hatzopoulos},\ and\ \citenamefont {Savvidis}}]{gao_2012}%
  \BibitemOpen
  \bibfield  {author} {\bibinfo {author} {\bibfnamefont {T.}~\bibnamefont
  {Gao}}, \bibinfo {author} {\bibfnamefont {P.~S.}\ \bibnamefont {Eldridge}},
  \bibinfo {author} {\bibfnamefont {T.~C.~H.}\ \bibnamefont {Liew}}, \bibinfo
  {author} {\bibfnamefont {S.~I.}\ \bibnamefont {Tsintzos}}, \bibinfo {author}
  {\bibfnamefont {G.}~\bibnamefont {Stavrinidis}}, \bibinfo {author}
  {\bibfnamefont {G.}~\bibnamefont {Deligeorgis}}, \bibinfo {author}
  {\bibfnamefont {Z.}~\bibnamefont {Hatzopoulos}}, \ and\ \bibinfo {author}
  {\bibfnamefont {P.~G.}\ \bibnamefont {Savvidis}},\ }\href {\doibase
  10.1103/PhysRevB.85.235102} {\bibfield  {journal} {\bibinfo  {journal}
  {Physical Review B}\ }\textbf {\bibinfo {volume} {85}},\ \bibinfo {pages}
  {235102} (\bibinfo {year} {2012})}\BibitemShut {NoStop}%
\bibitem [{\citenamefont {Kammann}\ \emph
  {et~al.}(2012{\natexlab{a}})\citenamefont {Kammann}, \citenamefont {Ohadi},
  \citenamefont {Maragkou}, \citenamefont {Kavokin},\ and\ \citenamefont
  {Lagoudakis}}]{kammann_crossover_2012}%
  \BibitemOpen
  \bibfield  {author} {\bibinfo {author} {\bibfnamefont {E.}~\bibnamefont
  {Kammann}}, \bibinfo {author} {\bibfnamefont {H.}~\bibnamefont {Ohadi}},
  \bibinfo {author} {\bibfnamefont {M.}~\bibnamefont {Maragkou}}, \bibinfo
  {author} {\bibfnamefont {A.~V.}\ \bibnamefont {Kavokin}}, \ and\ \bibinfo
  {author} {\bibfnamefont {P.~G.}\ \bibnamefont {Lagoudakis}},\ }\href
  {\doibase 10.1088/1367-2630/14/10/105003} {\bibfield  {journal} {\bibinfo
  {journal} {New Journal of Physics}\ }\textbf {\bibinfo {volume} {14}},\
  \bibinfo {pages} {105003} (\bibinfo {year} {2012}{\natexlab{a}})}\BibitemShut
  {NoStop}%
\bibitem [{\citenamefont {Deng}\ \emph {et~al.}(2003)\citenamefont {Deng},
  \citenamefont {Weihs}, \citenamefont {Snoke}, \citenamefont {Bloch},\ and\
  \citenamefont {Yamamoto}}]{deng_polariton_2003}%
  \BibitemOpen
  \bibfield  {author} {\bibinfo {author} {\bibfnamefont {H.}~\bibnamefont
  {Deng}}, \bibinfo {author} {\bibfnamefont {G.}~\bibnamefont {Weihs}},
  \bibinfo {author} {\bibfnamefont {D.}~\bibnamefont {Snoke}}, \bibinfo
  {author} {\bibfnamefont {J.}~\bibnamefont {Bloch}}, \ and\ \bibinfo {author}
  {\bibfnamefont {Y.}~\bibnamefont {Yamamoto}},\ }\href {\doibase
  10.1073/pnas.2634328100} {\bibfield  {journal} {\bibinfo  {journal}
  {Proceedings of the National Academy of Sciences}\ }\textbf {\bibinfo
  {volume} {100}},\ \bibinfo {pages} {15318} (\bibinfo {year} {2003})},\
  \bibinfo {note} {{PMID:} 14673089}\BibitemShut {NoStop}%
\bibitem [{\citenamefont {Tsotsis}\ \emph {et~al.}(2012)\citenamefont
  {Tsotsis}, \citenamefont {Eldridge}, \citenamefont {Gao}, \citenamefont
  {Tsintzos}, \citenamefont {Hatzopoulos},\ and\ \citenamefont
  {Savvidis}}]{tsotsis_2012}%
  \BibitemOpen
  \bibfield  {author} {\bibinfo {author} {\bibfnamefont {P.}~\bibnamefont
  {Tsotsis}}, \bibinfo {author} {\bibfnamefont {P.~S.}\ \bibnamefont
  {Eldridge}}, \bibinfo {author} {\bibfnamefont {T.}~\bibnamefont {Gao}},
  \bibinfo {author} {\bibfnamefont {S.~I.}\ \bibnamefont {Tsintzos}}, \bibinfo
  {author} {\bibfnamefont {Z.}~\bibnamefont {Hatzopoulos}}, \ and\ \bibinfo
  {author} {\bibfnamefont {P.~G.}\ \bibnamefont {Savvidis}},\ }\href {\doibase
  10.1088/1367-2630/14/2/023060} {\bibfield  {journal} {\bibinfo  {journal}
  {New Journal of Physics}\ }\textbf {\bibinfo {volume} {14}},\ \bibinfo
  {pages} {023060} (\bibinfo {year} {2012})}\BibitemShut {NoStop}%
\bibitem [{\citenamefont {Heller}\ \emph {et~al.}(1996)\citenamefont {Heller},
  \citenamefont {Filoramo}, \citenamefont {Roussignol},\ and\ \citenamefont
  {Bockelmann}}]{heller_direct_1996}%
  \BibitemOpen
  \bibfield  {author} {\bibinfo {author} {\bibfnamefont {W.}~\bibnamefont
  {Heller}}, \bibinfo {author} {\bibfnamefont {A.}~\bibnamefont {Filoramo}},
  \bibinfo {author} {\bibfnamefont {P.}~\bibnamefont {Roussignol}}, \ and\
  \bibinfo {author} {\bibfnamefont {U.}~\bibnamefont {Bockelmann}},\ }\href
  {\doibase 10.1016/0038-1101(95)00351-7} {\bibfield  {journal} {\bibinfo
  {journal} {Solid-State Electronics}\ }\textbf {\bibinfo {volume} {40}},\
  \bibinfo {pages} {725} (\bibinfo {year} {1996})}\BibitemShut {NoStop}%
\bibitem [{\citenamefont {Nagamune}\ \emph {et~al.}(1995)\citenamefont
  {Nagamune}, \citenamefont {Watabe}, \citenamefont {Sogawa},\ and\
  \citenamefont {Arakawa}}]{nagamune_onedimensional_1995}%
  \BibitemOpen
  \bibfield  {author} {\bibinfo {author} {\bibfnamefont {Y.}~\bibnamefont
  {Nagamune}}, \bibinfo {author} {\bibfnamefont {H.}~\bibnamefont {Watabe}},
  \bibinfo {author} {\bibfnamefont {F.}~\bibnamefont {Sogawa}}, \ and\ \bibinfo
  {author} {\bibfnamefont {Y.}~\bibnamefont {Arakawa}},\ }\href {\doibase
  doi:10.1063/1.114484} {\bibfield  {journal} {\bibinfo  {journal} {Applied
  Physics Letters}\ }\textbf {\bibinfo {volume} {67}},\ \bibinfo {pages} {1535}
  (\bibinfo {year} {1995})}\BibitemShut {NoStop}%
\bibitem [{\citenamefont {Gulia}\ \emph {et~al.}(1997)\citenamefont {Gulia},
  \citenamefont {Rossi}, \citenamefont {Molinari}, \citenamefont {Selbmann},\
  and\ \citenamefont {Lugli}}]{gulia_phonon-assisted_1997}%
  \BibitemOpen
  \bibfield  {author} {\bibinfo {author} {\bibfnamefont {M.}~\bibnamefont
  {Gulia}}, \bibinfo {author} {\bibfnamefont {F.}~\bibnamefont {Rossi}},
  \bibinfo {author} {\bibfnamefont {E.}~\bibnamefont {Molinari}}, \bibinfo
  {author} {\bibfnamefont {P.~E.}\ \bibnamefont {Selbmann}}, \ and\ \bibinfo
  {author} {\bibfnamefont {P.}~\bibnamefont {Lugli}},\ }\href {\doibase
  10.1103/PhysRevB.55.R16049} {\bibfield  {journal} {\bibinfo  {journal}
  {Physical Review B}\ }\textbf {\bibinfo {volume} {55}},\ \bibinfo {pages}
  {R16049} (\bibinfo {year} {1997})}\BibitemShut {NoStop}%
\bibitem [{\citenamefont {Kammann}\ \emph
  {et~al.}(2012{\natexlab{b}})\citenamefont {Kammann}, \citenamefont {Liew},
  \citenamefont {Ohadi}, \citenamefont {Cilibrizzi}, \citenamefont {Tsotsis},
  \citenamefont {Hatzopoulos}, \citenamefont {Savvidis}, \citenamefont
  {Kavokin},\ and\ \citenamefont {Lagoudakis}}]{kammann_nonlinear_2012}%
  \BibitemOpen
  \bibfield  {author} {\bibinfo {author} {\bibfnamefont {E.}~\bibnamefont
  {Kammann}}, \bibinfo {author} {\bibfnamefont {T.~C.~H.}\ \bibnamefont
  {Liew}}, \bibinfo {author} {\bibfnamefont {H.}~\bibnamefont {Ohadi}},
  \bibinfo {author} {\bibfnamefont {P.}~\bibnamefont {Cilibrizzi}}, \bibinfo
  {author} {\bibfnamefont {P.}~\bibnamefont {Tsotsis}}, \bibinfo {author}
  {\bibfnamefont {Z.}~\bibnamefont {Hatzopoulos}}, \bibinfo {author}
  {\bibfnamefont {P.~G.}\ \bibnamefont {Savvidis}}, \bibinfo {author}
  {\bibfnamefont {A.~V.}\ \bibnamefont {Kavokin}}, \ and\ \bibinfo {author}
  {\bibfnamefont {P.~G.}\ \bibnamefont {Lagoudakis}},\ }\href {\doibase
  10.1103/PhysRevLett.109.036404} {\bibfield  {journal} {\bibinfo  {journal}
  {Physical Review Letters}\ }\textbf {\bibinfo {volume} {109}},\ \bibinfo
  {pages} {036404} (\bibinfo {year} {2012}{\natexlab{b}})}\BibitemShut
  {NoStop}%
\bibitem [{\citenamefont {Savvidis}\ \emph {et~al.}(2000)\citenamefont
  {Savvidis}, \citenamefont {Baumberg}, \citenamefont {Stevenson},
  \citenamefont {Skolnick}, \citenamefont {Whittaker},\ and\ \citenamefont
  {Roberts}}]{savvidis_angle-resonant_2000}%
  \BibitemOpen
  \bibfield  {author} {\bibinfo {author} {\bibfnamefont {P.~G.}\ \bibnamefont
  {Savvidis}}, \bibinfo {author} {\bibfnamefont {J.~J.}\ \bibnamefont
  {Baumberg}}, \bibinfo {author} {\bibfnamefont {R.~M.}\ \bibnamefont
  {Stevenson}}, \bibinfo {author} {\bibfnamefont {M.~S.}\ \bibnamefont
  {Skolnick}}, \bibinfo {author} {\bibfnamefont {D.~M.}\ \bibnamefont
  {Whittaker}}, \ and\ \bibinfo {author} {\bibfnamefont {J.~S.}\ \bibnamefont
  {Roberts}},\ }\href {\doibase 10.1103/PhysRevLett.84.1547} {\bibfield
  {journal} {\bibinfo  {journal} {Physical Review Letters}\ }\textbf {\bibinfo
  {volume} {84}},\ \bibinfo {pages} {1547} (\bibinfo {year}
  {2000})}\BibitemShut {NoStop}%
\bibitem [{\citenamefont {Roumpos}\ \emph {et~al.}(2010)\citenamefont
  {Roumpos}, \citenamefont {Nitsche}, \citenamefont {Höfling}, \citenamefont
  {Forchel},\ and\ \citenamefont {Yamamoto}}]{roumpos_gain-induced_2010}%
  \BibitemOpen
  \bibfield  {author} {\bibinfo {author} {\bibfnamefont {G.}~\bibnamefont
  {Roumpos}}, \bibinfo {author} {\bibfnamefont {W.~H.}\ \bibnamefont
  {Nitsche}}, \bibinfo {author} {\bibfnamefont {S.}~\bibnamefont {Höfling}},
  \bibinfo {author} {\bibfnamefont {A.}~\bibnamefont {Forchel}}, \ and\
  \bibinfo {author} {\bibfnamefont {Y.}~\bibnamefont {Yamamoto}},\ }\href
  {\doibase 10.1103/PhysRevLett.104.126403} {\bibfield  {journal} {\bibinfo
  {journal} {Physical Review Letters}\ }\textbf {\bibinfo {volume} {104}},\
  \bibinfo {pages} {126403} (\bibinfo {year} {2010})}\BibitemShut {NoStop}%
\bibitem [{\citenamefont {Vishnevsky}\ \emph {et~al.}(2012)\citenamefont
  {Vishnevsky}, \citenamefont {Solnyshkov}, \citenamefont {Gippius},\ and\
  \citenamefont {Malpuech}}]{vishnevsky_multistability_2012}%
  \BibitemOpen
  \bibfield  {author} {\bibinfo {author} {\bibfnamefont {D.~V.}\ \bibnamefont
  {Vishnevsky}}, \bibinfo {author} {\bibfnamefont {D.~D.}\ \bibnamefont
  {Solnyshkov}}, \bibinfo {author} {\bibfnamefont {N.~A.}\ \bibnamefont
  {Gippius}}, \ and\ \bibinfo {author} {\bibfnamefont {G.}~\bibnamefont
  {Malpuech}},\ }\href {\doibase 10.1103/PhysRevB.85.155328} {\bibfield
  {journal} {\bibinfo  {journal} {Physical Review B}\ }\textbf {\bibinfo
  {volume} {85}},\ \bibinfo {pages} {155328} (\bibinfo {year}
  {2012})}\BibitemShut {NoStop}%
\bibitem [{\citenamefont {Love}\ \emph {et~al.}(2008)\citenamefont {Love},
  \citenamefont {Krizhanovskii}, \citenamefont {Whittaker}, \citenamefont
  {Bouchekioua}, \citenamefont {Sanvitto}, \citenamefont {Rizeiqi},
  \citenamefont {Bradley}, \citenamefont {Skolnick}, \citenamefont {Eastham},
  \citenamefont {André},\ and\ \citenamefont {Dang}}]{love_intrinsic_2008}%
  \BibitemOpen
  \bibfield  {author} {\bibinfo {author} {\bibfnamefont {A.~P.~D.}\
  \bibnamefont {Love}}, \bibinfo {author} {\bibfnamefont {D.~N.}\ \bibnamefont
  {Krizhanovskii}}, \bibinfo {author} {\bibfnamefont {D.~M.}\ \bibnamefont
  {Whittaker}}, \bibinfo {author} {\bibfnamefont {R.}~\bibnamefont
  {Bouchekioua}}, \bibinfo {author} {\bibfnamefont {D.}~\bibnamefont
  {Sanvitto}}, \bibinfo {author} {\bibfnamefont {S.~A.}\ \bibnamefont
  {Rizeiqi}}, \bibinfo {author} {\bibfnamefont {R.}~\bibnamefont {Bradley}},
  \bibinfo {author} {\bibfnamefont {M.~S.}\ \bibnamefont {Skolnick}}, \bibinfo
  {author} {\bibfnamefont {P.~R.}\ \bibnamefont {Eastham}}, \bibinfo {author}
  {\bibfnamefont {R.}~\bibnamefont {André}}, \ and\ \bibinfo {author}
  {\bibfnamefont {L.~S.}\ \bibnamefont {Dang}},\ }\href {\doibase
  10.1103/PhysRevLett.101.067404} {\bibfield  {journal} {\bibinfo  {journal}
  {Physical Review Letters}\ }\textbf {\bibinfo {volume} {101}},\ \bibinfo
  {pages} {067404} (\bibinfo {year} {2008})}\BibitemShut {NoStop}%
\bibitem [{\citenamefont {Porras}\ and\ \citenamefont
  {Tejedor}(2003)}]{porras_linewidth_2003}%
  \BibitemOpen
  \bibfield  {author} {\bibinfo {author} {\bibfnamefont {D.}~\bibnamefont
  {Porras}}\ and\ \bibinfo {author} {\bibfnamefont {C.}~\bibnamefont
  {Tejedor}},\ }\href {\doibase 10.1103/PhysRevB.67.161310} {\bibfield
  {journal} {\bibinfo  {journal} {Physical Review B}\ }\textbf {\bibinfo
  {volume} {67}},\ \bibinfo {pages} {161310} (\bibinfo {year}
  {2003})}\BibitemShut {NoStop}%
\bibitem [{\citenamefont {Holden}\ \emph {et~al.}(1997)\citenamefont {Holden},
  \citenamefont {Kennedy}, \citenamefont {Cameron}, \citenamefont {Riblet},\
  and\ \citenamefont {Miller}}]{holden_exciton_1997}%
  \BibitemOpen
  \bibfield  {author} {\bibinfo {author} {\bibfnamefont {T.~M.}\ \bibnamefont
  {Holden}}, \bibinfo {author} {\bibfnamefont {G.~T.}\ \bibnamefont {Kennedy}},
  \bibinfo {author} {\bibfnamefont {A.~R.}\ \bibnamefont {Cameron}}, \bibinfo
  {author} {\bibfnamefont {P.}~\bibnamefont {Riblet}}, \ and\ \bibinfo {author}
  {\bibfnamefont {A.}~\bibnamefont {Miller}},\ }\href {\doibase
  doi:10.1063/1.119694} {\bibfield  {journal} {\bibinfo  {journal} {Applied
  Physics Letters}\ }\textbf {\bibinfo {volume} {71}},\ \bibinfo {pages} {936}
  (\bibinfo {year} {1997})}\BibitemShut {NoStop}%
\end{thebibliography}
\end{document}